\documentstyle[preprint,revtex]{aps}
---------------------
\begin{document}
---------------------
\draft
\preprint{ISU-NP-93-15}
\preprint{LA-UR-94-784}

\vskip 0.5in

\begin{title}
J/$\Psi$ Suppression in Hadron-Nucleus Collisions
\end{title}

\vskip 0.4in

\author{Charles J. Benesh$^a$, Jianwei Qiu$^b$, and
James P. Vary$^{b,c}$}

\vskip 0.3in

\begin{instit}
$^a$Bonner Nuclear Laboratory, Rice University \\
Houston, TX 77251 and \\
Theoretical Division, MSB283, Los Alamos National
Laboratory, \\
Los Alamos, NM 87545 \\
\end{instit}

\vskip 0.2in

\begin{instit}
$^b$Department of Physics and Astronomy, Iowa State
University, \\
Ames, Iowa 50011 \\

\vskip 0.2in

$^c$Institute for Theoretical Physics, University of
Heidelberg, \\
Philosophenweg 19, D-69120 Heidelberg
\end{instit}

\newpage
--------------------

\begin{abstract}
	We examine the production of J/$\Psi$ mesons in
high energy
hadron-nucleus collisions using models that account for
both the
initial state modification of parton distributions in
nuclei
and the final state interaction of the produced $c\bar
c$ pairs.
We show that, at the energies of current fixed target
experiments,
J/$\Psi$ production through quark-antiquark annihilation
gives the largest contribution at $x_F > 0.5$, while
gluon-gluon
fusion dominates the production at smaller $x_F$.  The
observed
J/$\Psi$ suppression at large $x_F$ is
directly connected to nuclear shadowing in deeply
inelastic
scattering.  We find that a $6\sim 8$mb $c\bar
c$-nucleon cross
section is needed to explain the data on J/$\Psi$
suppression if
a simple incoherent multiple scattering formalism
is used for the final state interactions.
We then provide a more complete model to incorporate
the effects of final state multiple soft scatterings.
With only one parameter, this model can fit both the
normalization and A-dependence of the data for a variety
of nuclei.
\end{abstract}
--------------------


---------------------
\narrowtext
---------------------
\section{INTRODUCTION}

	Heavy quark production in high energy hadron
collisions
affords us a variety of insights into the underlying
dynamics of
the strong interaction. In proton-proton collisions, we
learn
about the interactions between the quarks and gluons
that make
up the colliding nucleons\cite{1}. In nucleus-nucleus
collisions,
we may find a signal for the existence of the
deconfining phase
of QCD, the quark-gluon plasma\cite{2}. In
proton-nucleus collisions,
the subject of this paper, we hope to gain information
about the
modification of the structure of the nucleon when it is
immersed in
the nuclear medium, and to learn, through the
final state interactions of the produced quarks with the
nucleons
that compose the nucleus,
of the processes through which quarks are transmuted
into the
observable hadrons we see experimentally. Recently, a
puzzle has arisen in
the pursuit of this knowledge.
Measurements of the A-dependence of charmonium
production in hadron-nucleus collisions at $\sqrt{s}$ of
20\cite{3} and
40 GeV\cite{4} have revealed an
unexpected suppression of the cross section
per nucleon at large values
of Feynman $x_F$. This suppression appears to be
in conflict with the predictions
of color transparency\cite{5}, where the final state
interactions of the
$c\bar c$ pair are expected to decrease with increasing
$x_F$ \cite{hue}.

In this paper, we demonstrate that a complete
description of nuclear J/$\Psi$
production requires proper treatment of both the
initial state modification of parton distributions in
the nucleus and
the final state interactions of the produced $c\bar c$
pairs. In particular,
we show that the existing data on J/$\Psi$ suppression
cannot be understood in terms of the subprocess of
gluon-gluon fusion
($gg\rightarrow c\bar c$) alone. At large values of
$x_F$, where the strong
suppression is observed, the
annihilation of light quark-antiquark pairs($q\bar
q\rightarrow c\bar
c$) dominates the J/$\Psi$ production cross section. In
light of this,
an accurate description of the modification of quark
distributions in the
nucleus is essential to any discussion of J/$\Psi$
production from nuclear
targets.
By incorporating both the EMC effect\cite{6} and
shadowing
at small
$x$\cite{NMC} into parton distributions in nuclei, and
using simple
models for the final state interactions of the produced
$c\bar c$
pair, we are able to explain the observed J/$\Psi$
suppression in
hadron-nucleus collisions.

In the next section, we  explain our
model for J/$\Psi$ production in hadronic collisions.
We
show that, at the energies of current fixed target
experiments,
quark-antiquark annihilation gives the largest
contribution to J/$\Psi$ production in hadronic
collisions at
$x_F > 0.5$, while gluon-gluon fusion dominates the
production
at smaller $x_F$. This result is shown to be easily
understood
since the probability for finding a valence quark at
large Bjorken $x$ in
the beam is much higher than the probability for finding
a gluon.
In section 3, we describe the approach we use to
determine the parton
distributions in a heavy nucleus.  To avoid unnecessary
model
dependence, we use high
statistics data on ratios of the structure functions to
fix
our parametrization of quark distributions in nuclei.
Since there is
no direct measurement of gluon distributions in nuclei,
we adopt two
different parametrizations, which represent a range of
uncertainties in the nuclear gluon distribution.
In section 4, we examine the effect of final state
interactions between the produced $c\bar c$ pair and
nuclear matter
on the J/$\Psi$ production cross section.
First, a simple incoherent multiple scattering (SIMS)
formalism
is used to calculate the final
state interactions with either a constant, energy
independent
J/$\Psi$-nucleon cross section, or
a cross section incorporating color transparency
effects (which reduce the final state interactions at
large $x_F$).
As an alternative to this picture, we consider a model
for final state
interactions which features multiple scattering between
produced
$c\bar c$ pairs and soft gluons as the pairs exit the
nucleus. As a
consequence of the multiple scatterings, the invariant
mass of the $c\bar c$
increases and, if the threshold for open charm
production is reached,
the pair hadronizes into open charm rather than
J/$\Psi$.
Finally, in section 5, we use the
parametrized nuclear parton distributions and models for
final state
interactions to calculate the
$x_F$-dependence of J/$\Psi$ suppression.
We discuss our numerical results, and their
uncertainties.
In particular, we examine the consequences of the
uncertainty in the
nuclear gluon distributions.

---------------------
\section{HADRONIC J/$\Psi$ PRODUCTION}

	In this section, we describe the model\cite{1}
we use to
calculate the J/$\Psi$ production cross section in
hadron-hadron collisions.
During the collision,
one parton (quark or gluon) from each hadron interacts
strongly to produce a $c\bar c$ pair. If the invariant
mass of
the pair is below the threshhold for open charm
production,
the pair can become a J/$\Psi$ particle through
radiation of
soft gluons.  Although the radiation of soft gluons is a
long-distance effect, which cannot be reliably
calculated within
perturbative QCD, it has been argued \cite{1} that such
radiation will affect only the overall normalization of
J/$\Psi$
production cross section.

The cross section for producing a J/$\Psi$ in collisions
of hadron $A$ and
hadron $B$ is given, up to an overall normalization, by
\cite{1}
\begin{equation}
\frac{d\sigma_{AB}}{dx_F} = \int_{4m_c^2}^{Q^2_{\rm
max}}
      \left(\frac{d\sigma_{AB}}{dQ^2dx_F}\right) dQ^2  ,
\label{eq1}
\end{equation}
where $x_F=x_1-x_2$, with $x_1$ and $x_2$ the momentum
fractions
carried by the partons originating in the beam and the
target,
respectively, and $Q^2$ is the invariant mass of the
produced $c\bar
c$ pair.  In Eq.~(\ref{eq1}), The integration limit
$4m_c^2$, with
$m_c\sim 1.5$~GeV, is the mass threshold for producing a
pair of charm
and anticharm quarks, and $Q^2_{\rm max}$ is the
threshold for
producing two open charm mesons
\begin{equation}
Q^2_{\rm max} = 4 m'^2
\label{eq2}
\end{equation}
with $m'\sim 1.85$~GeV \cite{1}.  The double
differential cross
section for producing $c\bar c$ pairs is given by
\begin{equation}
{d\sigma_{AB}\over dQ^2dx_F}={d\sigma^{q\bar
q}_{AB}\over dQ^2dx_F}
+{d\sigma^{c\bar c}_{AB}\over dQ^2dx_F}
+{d\sigma^{gg}_{AB}\over dQ^2dx_F},
\label{eq3}
\end{equation}
where $\sigma^{q\bar q}$ is the contribution from
annihilation of
light quarks, $q=u,d,s$, as shown in Fig.~1a;
$\sigma^{c\bar c}$ is
the contribution from intrinsic charm quarks, as shown
in
Fig.~1b; and $\sigma^{gg}$ is the contribution from
fusion of two
gluons, as shown in Fig.~1c.

The contribution of $\sigma^{c\bar c}$
should be much smaller than the other terms because the
very small
charmed sea distribution enters the
cross section quadratically \cite{SJB}.  The
contribution from the
other two terms are given by
\begin{equation}
{d\sigma^{q\bar q}_{AB}\over dQ^2dx_F}=\sum_{f=u,d,s}
{\hat{\sigma}^{q\bar q}(Q^2)\over Q^2}{x_1x_2\over
x_1+x_2}\times
\left[q_A^f(x_1)\bar q_B^f(x_2)+\bar q_A^f(x_1)
q_B^f(x_2)\right]
\label{eq4}
\end{equation}
for light quark annihilation, and
\begin{equation}
{d\sigma^{gg}_{AB}\over dQ^2dx_F}=
{\hat{\sigma}^{gg}(Q^2)\over Q^2}{x_1x_2\over x_1+x_2}
G_A(x_1)G_B(x_2),
\label{eq5}
\end{equation}
for gluon fusion.  Here, $q(x)$, $\bar q(x)$ and $G(x)$
signify the
quark, antiquark and gluon distributions, respectively,
in a hadron or
nucleus,
and the $Q^2$ dependence of the distributions is not
explicitly shown.

In Eqs.~(\ref{eq4}) and (\ref{eq5}), the momentum
fractions, $x_1$ and
$x_2$ can be expressed in terms of $Q^2$ and $x_F$ as,
\begin{equation}
x_1 ={1 \over 2}\left(\sqrt{x_F^2+4Q^2/s}+x_F\right) ,
\mbox{\hskip 0.5truein}
x_2 ={1 \over 2}\left(\sqrt{x_F^2+4Q^2/s}-x_F\right) ,
\label{eq5p}
\end{equation}
where $s$ is the invariant mass squared of the
hadron-hadron
collision.

The partonic cross sections appearing in
Eqs.~(\ref{eq4}) and
(\ref{eq5}) are given by\cite{1}
\begin{equation}
\hat{\sigma}^{q\bar q}(Q^2)={2\over 9}
{4\pi\alpha_s\over 3Q^2}(1+{1\over
2}\gamma)\sqrt{1-\gamma},
\label{eq6}
\end{equation}
and
\begin{eqnarray}
\hat{\sigma}^{gg}(Q^2)&=&{\pi\alpha_s\over 3Q^2}
\big{[}(1+\gamma+{1\over 16}\gamma^2)
\log({1+\sqrt{1-\gamma}\over
1-\sqrt{1-\gamma}})\nonumber\\
& &\,\,\,-({7\over 4}+{31\over
16}\gamma)\sqrt{1-\gamma}\,\big{]},
\label{eq7}
\end{eqnarray}
where $\alpha_s$ is the QCD running coupling constant,
and $\gamma=4m_c^2/Q^2$.

To understand the relative size of the contributions
from light quark
annihilation and gluon fusion, we calculated the
fraction of the total
$c\bar c$ production cross section due to each
subprocess in a proton-deuteron
collision, using the leading order parton distributions
of
Morfin and Tung \cite{13} for the proton, and the
isospin average of
the same parton distributions for the deuteron. The
result, shown in
Fig. 2, is very similar to what one obtains for a
heavier nuclear target.
(Note that the sum of the two fractions is unity
since we have neglected the contribution of intrinsic
charm.)
The fraction of the cross section due to gluon fusion,
indicated by
the solid line in Fig. 2, dominates the cross section
for $x_F<0.5$,
while $q\bar q$ annihilation, shown by the dashed line,
is largest at
higher $x_F$. This result is easily understood, since
large $x_F$
requires a beam parton at large $x_1$, where the gluon
distribution of the proton is very small relative to the
valence quark
distribution.
Consequently, the J/$\Psi$ production cross section at
large $x_F$ is
 very sensitive to the sea quark distributions of the
target. In particular,
the shadowing of sea quark distributions in the nucleus
will play a
crucial role in determining
the nuclear dependence of J/$\Psi$ production cross
sections at large
$x_F$.

---------------------
\section{PARTON DISTRIBUTIONS IN NUCLEI}

We now turn to a description of the modifications of the
initial
state parton distributions in the target nucleus.
A variety of approaches to this question exist in the
literature\cite{9}. In order to avoid unnecessary model
dependence, we
refrain from using any particular model for the
medium-modified quark
distributions, the EMC effect \cite{6}, and develop
instead a
parametrization of the measured ratio of structure
functions
in lepton-nucleus deeply inelastic scattering (DIS),
$R(x)$,
based on simple physics considerations.  For the gluon
distributions, we adopt two parametrizations in order to
gauge the sensitivity
of our results to model dependent assumptions.

For quark distributions in a nucleus, we begin by
dividing the
quark momentum fraction, $x$, into three regions,
which we shall call the ``Fermi'' region ($x>0.6$),
the ``Old EMC'' region ($x_0 <x<0.6$), with $x_0\sim
0.15$,
and the ``Shadowing'' region ($x<x_0)$, respectively.
Since $0<x_F<0.7$ for existing data on J/$\Psi$
production,
the momentum fraction $x_2$ of partons
from the nuclear target will be limited to $x_2<0.3$.
Therefore, the Fermi region is not relevant to the
present discussion
and we give no parametrizion for $R(x)$ in this paper.
In the Old EMC region, where various models espouse
different physical mechanisms, data shows a clear
$x$-dependence,
and a weak dependence on atomic weight $A$.
Since we are mainly concerned with $x\leq 0.3$, we
ignore the
A-dependence and parametrize the ratio of nuclear and
deuteron
quark distributions, $R^q_{A/D}(x)$,
by a simple linear function
\begin{equation}
R^q_{A/D}(x)=\alpha_3 - \alpha_4\ x
\mbox{\hskip 0.8truein for }  x_0\leq x \leq 0.6,
\label{eq16}
\end{equation}
where $x_0\sim 0.15$, and $\alpha_3$ and $\alpha_4$ are
parameters
given in Table I.

As discussed in the previous section, accuracy of
nuclear quark distributions
in the Shadowing region is
crucial to understanding the existing data on J/$\Psi$
production at
large $x_F$.
Fortunately, shadowing in effective nuclear quark
distributions has been measured through DIS.
In principle, we can choose any function with enough
parameters for
$R^q_{A/D}(x)$, and use the data to fix those
parameters.
However, we prefer to invoke some physics arguments to
motivate a
functional form.

It has been demonstrated by Fermilab experiment E665
\cite{E665} that
the ratio of structure functions in DIS saturates as
$x\rightarrow 0$.
Consequently, $R^q_{A/D}(x)$ should also saturate as
$x\rightarrow 0$,
and quark distributions in a large nucleus should have
the same
small-$x$ behavior as those in the deuteron.
In the parton recombination picture, shadowing is
due to the recombination of soft partons \cite{12},
and the effective nuclear quark distribution can be
parametrized
as \cite{JQ}
\begin{equation}
q_A^f(x) \propto q_0^f(x)\left[1 - k_q A^{\alpha_1}
         \left({1\over x}-{1\over x_0}\right)\right]
\label{eq17}
\end{equation}
for quark flavor $f$.  In Eq.~(\ref{eq17}),
$q_0^f(x)$ represents a ``primordial'' quark
distribution without any
recombination effect.  Thus, $q_0^f(x)$ should not be
considered as
a proton quark distribution which acquires its own
recombination
effect \cite{12}.
The ratio of quark distributions of two nuclei with
atomic weights,
$A_1$ and $A_2$, respectively, is defined as
\begin{equation}
R^f_{A_1/A_2}(x) = \frac{q_{A_1}^f(x)}{q_{A_2}^f(x)},
\label{eq18}
\end{equation}
where all effective nuclear parton distributions are
normalized by
corresponding atomic weights.  Combining
Eqs.~(\ref{eq17}) and
(\ref{eq18}), we choose the following
parametrization for $R^q_{A/D}(x)$ in the region of our
interest,
\begin{equation}
R^q_{A/D}(x) = \left\{\begin{array}{ll}
R^q_{A/D}(x_0)
     \frac{{\normalsize 1/\left[1 + k_q A^{\alpha_1}

\left({1/x}-{1/x_0}\right)\right]}}
          {{\normalsize 1/\left[1 + k_q \alpha_2

\left({1/x}-{1/x_0}\right)\right]}}
     & \mbox{\hskip 0.5truein} x<x_0 \\
     &                               \\
\alpha_3 - \alpha_4 x
     & \mbox{\hskip 0.5truein} x_0\leq x \leq 0.6
\end{array} \right.
\label{eq19}
\end{equation}
where we have used the approximation $1-\delta\sim
1/(1+\delta)$,
if $\delta$ is small, to smooth the limit when $x$
approaches to zero.
In addition, we introduce an extra fitting parameter
$\alpha_2$ to
replace $A^{\alpha_1}$ for the deuteron.
In Eq.~(\ref{eq19}), $R^q_{A/D}(x_0)= \alpha_3 -
\alpha_4 x_0$.

Using the high statistics data from NMC \cite{NMC}, we
can fix all
parameters: $k_q$ and $\alpha_i$, $i=1,2,3$ and 4.
The resulting values are given in table I.
In Fig.~3, we compare the NMC data with our
parametrization given in
Eq.~(\ref{eq19}).  We demonstrate the saturation when
$x\rightarrow 0$ in Fig.~4, where we compare our
parametrization
with the smaller $x$,
but, lower statistics data from E665 \cite{E665}.
Note, our fit was obtained by fitting the NMC data in
Fig.~3 only.
It is clear that our parametrization gives a good
description of the
ratios of nuclear and deuteron quark distributions in
the
small-$x$ region for A$\geq 4$.

Since there are no direct measurements
of gluon distributions in a nucleus,
we adopt two parametrizations.
In SET 1, we assume that the ratio of gluon
distributions,
$R^g_{A/D}(x)$, is very similar to $R^q_{A/D}(x)$.  In
the parton
recombination picture, such an assumption overestimates
gluon
shadowing \cite{JQ}.
The other parametrization, SET 2,  approximately
reproduces the two calculations on perturbative
shadowing in
Refs.~\cite{7} and \cite{eqw}, and serves as a lower
limit on
gluon shadowing.  The parametrized $R^g_{A/D}(x)$ for
both SET 1 and
SET 2 is given by
\begin{equation}
R^g_{A/D}(x)= a_A \left\{\begin{array}{ll}
\frac{1/\left[1 + k_g
A^{\alpha_1}\left({1/x}-{1/x_0}\right)

/x^{\alpha_3}\right]}
     {1/\left[1 + k_g \alpha_2
\left({1/x}-{1/x_0}\right)

/x^{\alpha_3}\right]}
         & \mbox{\hskip 0.5truein} x<x_0 \\
         &                \\
1.0
         & \mbox{\hskip 0.5truein} x_0\leq x \end{array}
\right.
\label{eq20}
\end{equation}
where $a_A$ is determined by momentum conservation,
\begin{equation}
a_A = \frac{f_0 (1+\epsilon_A)}
           {\int_0^{x_0} dx xG(x) \bar{R}^g_{A/D}(x) +
            \int_{x_0}^1 dx xG(x)},
\label{eq21}
\end{equation}
$f_0=\int_0^1 dx xG(x)$ is the fraction of momentum
carried by
gluons, $\epsilon_A$ is the momentum transfer from
quarks to gluons in
a nucleus, and $\bar{R}^g_{A/D}(x)=R^g_{A/D}(x)/a_A$.

In our numerical calculations, we use the parton
distributions of
Morfin and Tung \cite{13} for the proton and the
neutron,
and choose $\epsilon_A \sim 0.0$.  The two
sets of parameters: $k_g$ and $\alpha_i$, $i=1,2$, and
3, are given in
Table II.  For comparison, we plot the two choices for
$R^g_{A/D}(x)$
 in Fig.~5.

---------------------
\section{FINAL STATE INTERACTIONS}

In this section, we  carefully examine models for the
final state interactions of the produced $c\bar c$ pairs
passing
through the nucleus.  First, we use the SIMS formalism
with
and without a model of the color transparency mechanism
(subsection
\ref{ss4.1}).  The net effect of the final state
interactions in this
approach can be expressed as a multiplicative factor
$A_{eff}(Q^2,x_F)$ (given below in Eq.~(\ref{eq12}))
applied to the right-hand-side of Eqs.~(\ref{eq4}) and
(\ref{eq5}).
We then introduce in subsection \ref{ss4.2} a new model
to
incorporate the effect of final state multiple soft
scattering.
In this model, we account for the net inelasticity of
the
multiple scatterings by adjusting the upper limit of the
$Q^2$-integration (open charm threshold) in
Eq.~(\ref{eq1}) to reflect
the increase of the invariant mass of a colored $c\bar
c$ pair as it
exits the nucleus.

\subsection{SIMS formalism with and without color
transparency}
\label{ss4.1}

In the usual treatment of color transparency($CT$)
\cite{5,hue}
the $c\bar c$ pair is assumed to
have been produced in a color singlet state
of small spatial extent of order $1/m_c$ or less. As the
pair exits the
nucleus, it expands.  But at
high energies the system escapes well before any
significant expansion
occurs, and can be thought to exist in a small size
state for
the entire time it is inside the target. As a result,
the color
singlet pair has a small color dipole moment
while it traverses the nucleus,
and the probability of interactions breaking the pair
into open charm
is small. Assuming that the pair has an initial size of
order $R_0(Q)$,
where $Q$ is the invariant mass of the pair, the
interaction
cross section can be modeled by \cite{5},
\begin{equation}
 \sigma_{CT}(z)=\sigma_0\Big{(}{r(z)\over
R_{J/\Psi}}\Big{)}^2,
\label{eq8}
\end{equation}
where $\sigma_0$ is the J/$\Psi$-nucleon cross section,
whose value
will be discussed in section~5.  In Eq.~(\ref{eq8}),
$R_{{\rm J}/\Psi}$ is the J/$\Psi$ radius, and $r(z)$ is
given by
\begin{equation}
r(z)=\min \Big{(}R_{{\rm J}/\Psi}, (R_0(Q) +z {2m_N
v_\perp
\over \sqrt{s} v_\parallel})\Big{)},
\label{eq9}
\end{equation}
where $s$ is the invariant mass squared
of the nucleon-nucleon collision, and $R_0(Q)$
is the initial size of the $c\bar c$ pair,
 $z$ is the distance traveled from the
 $c \bar c$ creation point in the lab frame,
\begin{equation}
v_\parallel ={x_F\over\sqrt{x_F^2+4Q^2/s}}
\label{eq10}
\end{equation}
is the longitudinal speed of the pair in the
nucleon-nucleon center of mass
frame,
and $v_\perp$ is the transverse speed of the pair in the
nucleon-nucleon
center of mass frame. The value of $v_\perp$
governs how fast the $c\bar c$ system
expands to the size of J/$\Psi$. Our choice
is $v_\perp\sim1$ based on the
uncertainty principle argument coupled with
the nearly point like creation of the
$c\bar c$ pair. Other expansion models can
also be examined \cite{5} amounting to
slower expansion rates. We have not performed
an exhaustive examination of various
$CT$ models since we find enough freedom
within the $CT$ model we use to fit the data, provided
$\sigma_0$ is
taken sufficiently large.
The probability that a pair, created at the point $\vec
r=({\vec r_{\perp}},z_0)$, escapes the nucleus with
density
distribution, $\rho(\vec r)$, without interacting is
given by
\begin{equation}
P_{esc}(\vec r,Q^2,x_F) =
{\rm exp}\left[-\int_{z_0}^\infty dz^\prime \rho({\vec
r}^\prime)
\sigma_{CT}(z^\prime-z_0)\right].
\label{eq11}
\end{equation}
The total fraction of escaping J/$\Psi$'s is
obtained by averaging over
the creation points weighted by the nuclear density,
\begin{equation}
{A_{eff}(Q^2,x_F)\over A}={1\over A}\int d^3r \rho(\vec
r)
P_{esc}(\vec r,Q^2,x_F),
\label{eq12}
\end{equation}
where the dependence on $Q^2$ and $x_F$ is through
Eqs.~(\ref{eq9}) and
(\ref{eq10}).

\subsection{Multiple soft scattering of the colored
$c\bar c$ pair}
\label{ss4.2}

Observed anomalous nuclear enhancement in the
acoplanarity of
two-jet production\cite{corc} tells us that a colored
parton (quark or gluon)
experiences multiple soft scatterings when it exits the
target
nucleus \cite{lqs}.  In our model of J/$\Psi$
production,
the $c\bar c$ pair produced through partonic hard
scattering is not a color singlet state.
Such a colored $c\bar c$ pair should
also experience multiple soft scatterings when it passes
through
nuclear matter.
These multiple soft scatterings will increase the
relative transverse
momentum between the $c$ and $\bar c$, and consequently,
increase the
invariant mass of the $c\bar c$ pair.
Some  pairs will gain
enough mass to be pushed over the threshold
and  become two open charm mesons, and the J/$\Psi$
production cross section
will be reduced in comparison with nucleon targets. In
larger nuclei,
the pair will undergo more soft scatterings, and the
reduction in the cross
section for J/$\Psi$ production will be even greater.
In this model of J/$\Psi$ suppression, we account for
the final state
interactions by adjusting
the upper limit of $Q^2$-integration in Eq.~(\ref{eq1})
as
\begin{equation}
Q^2_{\rm max}\rightarrow Q^2_{\rm max}-E^2_{\rm ms},
\label{eq7p}
\end{equation}
where $E^2_{\rm ms}$ is the increase in the invariant
mass squared
 of the colored
$c\bar c$ pair as it exits the nucleus.

	To understand the A dependence of $E^2_{ms}$,
consider the
following simple model.  A $c\bar c$ pair of invariant
mass Q,
 interacts, in its rest frame, with the
nucleus by emitting and absorbing soft gluons. The time
scale for emitting
a gluon is of the same order of magnitude as the time
scale for hadronization,
and therefore the pair cannot lower its internal energy
while it is in the
nucleus. Absorption, on the other hand, can occur since
the flux of gluons
incident on the pair is enormous. When the pair absorbs
a gluon of momentum
 $k=(k,\vec k)$, its invariant mass increases by
\begin{equation}
E_1^2 = (P+k)^2-P^2 = 2k Q.
\label{eqe1}
\end{equation}
After $n$-fold soft scatterings, the gain of the
invariant mass will
be
\begin{equation}
E_n^2 \approx 2 n \langle k \rangle Q ,
\label{eqe2}
\end{equation}
if $n \langle k \rangle \ll Q$
with $\langle k \rangle$ an average momentum of the
soft gluons.
In a target nucleus, the number
of scatterings, $n$, should be proportional to the
nuclear size, which
is proportional to A$^{1/3}$.
We have no independent experimental input for $\langle k
\rangle$
although some information is available for the $\langle
k_T^2 \rangle$
of two-jet momentum imbalance \cite{lqs,corc}.  It is
beyond the scope
of the present effort to develop a detailed model
linking two-jet
production with the experiment we address here.
Therefore, we adopt a one-parameter
expression for the square of energy gained by the
colored
$c\bar c$ pair due to multiple soft scatterings
\begin{equation}
E^2_{\rm ms} = \epsilon A^{1/3}
\label{eq14}
\end{equation}
where the value of $\epsilon$ will be determined by
fitting the data
on J/$\Psi$ suppression.  As we will show in section 5,
with this
one-parameter expression, our model for J/$\Psi$
suppression can fit
both the normalization and A-dependence of the data for
a variety of
nuclei.

---------------------
\section{NUMERICAL RESULTS AND DISCUSSIONS}

Having presented the cross section for J/$\Psi$
production,
Eq.~({\ref{eq1}), the effective nuclear parton
distributions in
section 3, and the models to calculate the final state
interactions
from the previous section, we now turn to the
presentation and
discussion of numerical results.

We define the ratios of cross sections for J/$\Psi$
production,
$R^{{\rm J}/\Psi}_{A/D}(x_F)$, from nuclear (A) and
Deuterium (D)
targets with a proton beam as
\begin{equation}
R^{{\rm
J}/\Psi}_{A/D}(x_F)=\frac{1}{A}\frac{d\sigma_{pA}}{dx_F}
\left/
\frac{1}{2}\frac{d\sigma_{pD}}{dx_F}\right.
\label{eq22}
\end{equation}
where $d\sigma/dx_F$ is given in Eq.~(\ref{eq1}).
To evaluate the numerator we form the quark [gluon]
distributions in
the nucleus by multiplying Eq.~(\ref{eq19})
[Eq.~(\ref{eq20})] by the
simple isospin average of the quark [gluon]
distributions for a free
nucleon.
The denominator is formed by the isospin average of the
proton and
neutron contribution.
The mass threshold in Eq.~(\ref{eq1}) is chosen as:
$m_c=1.5$~GeV, and
$m'=1.85$~GeV \cite{1}.  The
ratio $R^{{\rm J}/\Psi}$ is not very sensitive to the
actual value of
the mass threshold, $m'$, as long as $m'\sim 1.85$~GeV.

In Fig.~6, we plot the ratios of cross sections of
J/$\Psi$ production
for several nuclei, along with Fermilab E772 data at
$\sqrt{s}=40$ GeV
\cite{4}.  In these figures, the SET~2 gluon
distribution in nuclei
and the SIMS formalism of subsection \ref{ss4.1} with a
constant
$\sigma_0$, defined in Eq.~(\ref{eq8}), were used. The
solid curve
indicates the ratio obtained if no final state
interactions are included
in the model.
The long dashed curve is the ratio assuming a
J/$\Psi$-nucleon cross section
of 3mb \cite{3}. Clearly, this curve does not provide
enough suppression to reproduce the data.
On the other hand, the $c\bar c$ pair produced at
short-distance through
the hard scattering diagrams in Fig.~1 does not have the
quantum numbers
of the J/$\Psi$ meson. Because the s-channel diagrams in
Fig.~1 dominate
the production, the pair will have the same quantum
numbers as a gluon.
Assuming that all other factors are equal,
we parametrize the colored $c\bar c$-nucleon cross
section to be the
J/$\Psi$-nucleon cross section multiplied by a color
enhancement factor.
In the leading pole approximation \cite{lqs}, soft gluon
interactions
with the $c\bar c$ pairs will give a color enhancement
factor
$C_g=N_c=3$ with $N_c$ being the number of colors. The
resulting
J/$\Psi$ production ratio, given by the dot-dashed curve
in Fig. 6,
lies below the data, indicating that the $c\bar
c$-nucleon cross section is
slightly too large. A reduction of this cross section is
easily understood
since the produced $c\bar c$ pair does not have the size
of a
J/$\Psi$ meson, as we have assumed.
Without a detailed $\chi^2$ fit,
we find that, with a $6-8$mb effective $c\bar c$-nucleon
cross section, we can obtain a visually good description
to the data for a variety of nuclei.

$CT$ will reduce final state nuclear effects.  With
the model of $CT$ presented in the last section, and
with
the initial size of the produced $c\bar c$ pair
$R_0(Q)=0$, we find that the ratios will be about 10\%
higher than
the dashed and dot-dashed curves which we have plotted
in Fig.~6.
However, the final state suppression
in the $CT$ model is very sensitive to the choice of
$R_0(Q)$.
Indeed, if
one assumes $R_0(Q)=1/Q$, curves with the $CT$ model
and that with constant J/$\Psi$-nucleon cross section
are virtually
indistinguishable.  Thus, there seems to be ample
freedom for $CT$ models,
possibly in combination with other effects,
to describe these data. We therefore
arrive at the conclusion that the J/$\Psi$
production data neither prove nor
contradict $CT$. It is worth
noting here that recent results from $(e,e'p)$ reactions
\cite{14}, also fail to provide a definitive
test of $CT$ effects.

In Fig.~7, we plot the same data from Fermilab E772 on
the
ratios of J/$\Psi$ production cross sections, but, with
the new
model for the final state interactions from subsection
\ref{ss4.2}.
In this model, the collective effect of
final state soft scatterings was implemented through
a nucleus-dependent upper limit of $Q^2$-integration as
\begin{equation}
\frac{d\sigma_{pA}}{dx_F} = \int_{4m_c^2}^{4m'^2-E_{\rm
ms}^2}
      \left(\frac{d\sigma_{pA}}{dQ^2dx_F}\right) dQ^2  ,
\label{eq23}
\end{equation}
where $E_{\rm ms}^2$ is given in Eq.(\ref{eq14}).  The
only parameter,
$\epsilon$, was chosen to be $\epsilon=0.3$ GeV$^2$
to give a reasonable fit to the data as shown in Fig.~7.
The solid and dashed lines correspond
to SET~2 and SET~1 effective nuclear gluon
distributions,
respectively, which
represent a range of uncertainty in the gluon
distributions
inside a large nucleus.  The dotted lines in Fig.~7
include only
the initial state nuclear effect in parton
distributions.
It is clear that our model of J/$\Psi$ suppression
can give a good fit to both the normalization and
A-dependence
of the data for a variety of nuclear targets, with just
one free
parameter, $\epsilon$.

In addition to the final state multiple soft scatterings
discussed
above, hard scattering between the $c\bar c$ pair and
nucleons inside
a target nucleus can also result in some suppression of
J/$\Psi$
production.  To estimate the effect of hard scattering,
we use the
SIMS formalism with the $c\bar c$-nucleon cross section
$\sigma_0$ (defined in Eq.~(15)) to be
$\sigma_0|_{\rm hard}=1/Q^2$.
This is consistent with the argument, given by
Brodsky and Mueller\cite{BM}, that the J/$\Psi$ meson
will be formed far outside of the nucleus in
hadron-nucleus collisions.
By combining the suppressions due to both soft and hard
final state
interactions, we plot our results in Fig.~8.  The data
and the solid
and dotted lines are the same as those in Fig.~7.  The
dashed lines
include both soft and hard scattering.  This result is
consistent with
general expectation that soft scattering dominates.

Having dealt successfully with the data at $\sqrt{s}$=40
GeV,
we also plot the comparison of our results with the
NA3 data at 20 GeV \cite{3}.  The ratio of the J/$\Psi$
production cross sections on deuterium and heavy Pt
target is shown
in Fig.~9.  Note that, following the presentation of the
experimental
results, the ratio plotted is the reciprocal of the
ratio shown in
Figs.~6-8.  The curves are calculated for a proton beam.
We used the SET 2 gluon distribution
in nuclei, and the same value of $\epsilon$ as that used
in Fig.~7.
Again, the dashed curve includes only the initial state
effect.
The significant differences between the data at 20 and
40 GeV has been
discussed by many authors \cite{4,hoy}.  It is our
impression that our
results are in reasonable accord with the data at both
20 GeV and 40
GeV.

In conclusion, we have demonstrated in this paper that
it is possible
to understand the observed strong nuclear suppression in
J/$\Psi$
production in hadron-nucleus collisions.
By combining both initial and final state nuclear
effects,
we are able to fit the data on a variety of nuclear
targets.  We have presented two models for the final
state
interactions between the $c\bar c$ pairs and target
nuclear matter.
Although these two models seem to be very different,
they are both based on the underlying picture of
multiple scattering in nuclear matter.
In the SIMS model we include multiple independent
(incoherent)
scattering which simply attenuate the $c\bar c$ pair.
The second model
incorporates both the independent hard scatterings
(which is a small effect) and multiple soft scattering.
The accumulated effect of the soft scatterings
is shown to be the dominant contribution to J/$\Psi$
suppression
in this picture. In either picture,
J/$\Psi$ suppression, in principle, is
accompanied with
enhancement of open charm production.  However, such
enhancement will
not have any significant experimental consequence
because the cross
section for
the J/$\Psi$ production is about three orders of
magnitude smaller than
that of open charm production.

Finally, it should be noted that it is not our purpose
here to produce a
detailed fit to get the best value of $\epsilon$ in our
model, and we
are certainly not claiming that the final state soft
scattering is the
only source for
the suppression observed in J/$\Psi$ production.  On the
contrary, we
continue to explore other possible effects responsible
for the
suppression.  Nevertheless, from the consistency between
the model and
the observed data, we believe that parton shadowing and
final state
multiple scattering effects are likely to be the most
important
sources of the J/$\Psi$ suppression observed in
hadron-nucleus collisions.

\newpage
------------------
\acknowledgements

We are pleased to acknowledge stimulating
discussions with J. H\"ufner, A.H. Mueller, E. Quack, M.
Simbel and G.
Sterman.  This work was supported in part by the U.S.
Department of
Energy, Division of High Energy and Nuclear
Physics, ER-23 and Grant Nos.
DE-FG02-87ER40371 and DE-FG02-92ER40370, and
by Texas National Research Laboratory Commission.
One of us (JPV) wishes
to acknowledge support from the Alexander
von Humboldt-Stiftung.

---------------------

--------------------
--------------------
\figure{Feynman diagrams for production of a $c\bar c$
pair
through (a) annihilation of light quarks, (b) intrinsic
charm quarks, and (c) fusion of two gluons.}

\figure{Ratio of quark and gluon fusion cross sections
in proton-deuteron
collisions to their sum as a function of $x_F$.  The
solid line is for
the quark contribution, and the dashed line for the
gluon contribution.}

\figure{Comparison of the NMC data \cite{NMC} for the
ratio of structure
functions for other nuclear targets to Deuterium along
with the
parametrization given in Eq.(\ref{eq19}).  }

\figure{Comparison of the E665 data \cite{E665} for the
ratio of
structure functions for Xe to Deuterium along with the
parametrization
given in Eq.(\ref{eq19}). }

\figure{Ratio of gluon distributions for various nuclei
to Deuterium.
The parametrization is given in Eq.~(\ref{eq20}).  The
dashed and solid
lines correspond to SET~1 and SET~2 parameter values
given in Table II.}

\figure{Ratio of J/$\Psi$ production cross section for
different nuclear targets to Deuterium at $\sqrt{s}$=40
GeV.
Data are from E772 [4].  The solid lines include only
the initial
state nuclear effects in the parton distributions.
The dashed and dot-dashed curves include both
initial and final state nuclear effects with final state
interactions
calculated with the SIMS formalism (subsection
\ref{ss4.1}) with a
$c\bar c$-nucleon cross section of 3 and 9 mb,
respectively.  SET~2
gluon distributions in nuclei are used.}

\figure{Ratio of J/$\Psi$ production cross sections for
different nuclear targets to Deuterium at $\sqrt{s}$=40
GeV.
Data are from E772 [4].
The dotted lines include only initial state nuclear
effects in the
parton distributions.  The solid and dashed curves
include both
initial and final state nuclear effects with final state
interactions
calculated in our multiple soft scattering model of
subsection
\ref{ss4.2} with $\epsilon = 0.3$~GeV$^2$.
The solid and dashed curves correspond to the SET~2 and
SET~1 gluon
distributions, respectively.}

\figure{Ratio of J/$\Psi$ production cross section for
different nuclear targets to Deuterium at $\sqrt{s}$=40
GeV.
Data are from E772 [4].
The dotted and solid lines are the same as those in
Fig.~7.  The
dashed lines include also hard scattering between $c\bar
c$ pairs and
nucleons inside target nuclei as described in the text.
The $c\bar c$-nucleon hard-scattering cross section is
$\sigma_0=1/Q^2$ with $Q^2$ defined as
the invariant mass of the $c\bar c$ pair.}

\figure{Ratio of $J/\Psi$ production cross section for
Deuterium to Pt at  $\sqrt{s}$=20 GeV.
The curves represent a calculation for a proton beam.
The dashed curve includes only initial state nuclear
effects in parton
distributions.  The solid curve includes both
initial and final state nuclear effects with final state
interactions
calculated in our multiple soft scattering model with
$\epsilon = 0.3$~GeV$^2$.  Also shown are the NA3 proton
and
pion data \cite{3}. Note that the ratio plotted is
the inverse of the ratio presented in Figs.~6-8.}

---------------------
\mediumtext
\begin{table}
\setdec 0.0000
\caption{Values of parameters for $R^q_{A/D}$ defined in
Eq.~(\ref{eq19}).}
\begin{tabular}{cccccc}
$x_0$ & $k_q$ & $\alpha_1$ & $\alpha_2$ &
                $\alpha_3$ & $\alpha_4$    \\
\tableline
0.1400 & 0.0098 & 0.1000 & 1.0163 & 1.0551 & 0.2322 \\
\end{tabular}
\end{table}

\begin{table}
\setdec 0.0000
\caption{Values of parameters for $R^g_{A/D}$ defined in
Eq.~(\ref{eq20}).}
\begin{tabular}{cccccc}
       & $x_0$ & $k_g$ & $\alpha_1$ & $\alpha_2$ &
$\alpha_3$
      \\
\tableline
 SET 1 & 0.1400 & 0.0098 & 0.1000 & 1.0163 & 0.0000
      \\
\tableline
 SET 2 & 0.05   & 0.010   & 0.100  & 1.016 & -0.300
      \\
\end{tabular}
\end{table}

---------------------
\end{document}